\documentclass[twocolumn]{aastex631}
\usepackage{sidecap}
\usepackage{amsmath}
\usepackage[normalem]{ulem}
\usepackage{tabularx}
\DeclareUnicodeCharacter{0301}{\'{e}}

\begin{document}

\title{Inferring coupling strengths of mixed-mode oscillations in red-giant stars using deep learning}

\author{Siddharth Dhanpal}
\email{dhanpal.siddharth@gmail.com}
\affiliation{Department of Astronomy and Astrophysics, Tata Institute of Fundamental Research, Mumbai, 400005, India}

\author{Othman Benomar}
\affiliation{ Center for Space Science, NYUAD Institute, New York University Abu Dhabi, PO Box 129188, Abu Dhabi, UAE}
\affiliation{Division of Solar and Plasma Astrophysics, NAOJ, Mitaka, Tokyo, Japan}

\author{Shravan Hanasoge}
\affiliation{Department of Astronomy and Astrophysics, Tata Institute of Fundamental Research, Mumbai, 400005, India}
\affiliation{ Center for Space Science, NYUAD Institute, New York University Abu Dhabi, PO Box 129188, Abu Dhabi, UAE}

\author{Masao Takata}
\affiliation{Department of Astronomy, School of Science, The University of Tokyo, 7-3-1 Hongo, Bunkyo-ku 113-0033, Tokyo, Japan}

\author{Subrata Panda}
\affiliation{Department of Astronomy and Astrophysics, Tata Institute of Fundamental Research, Mumbai, 400005, India}

\author{Abhisek Kundu}
\affiliation{Parallel Computing Lab, Intel Labs, Bangalore, India}

\begin{abstract}
Asteroseismology is a powerful tool that may be applied to shed light on stellar interiors and stellar evolution. Mixed modes, behaving as acoustic waves in the envelope and buoyancy modes in the core, are remarkable because they allow for probing the radiative cores and evanescent zones of red-giant stars. Here, we have developed a neural network that can accurately infer the coupling strength, a parameter related to the size of the evanescent zone, of solar-like stars in $\sim$5 milliseconds. In comparison with existing methods, we found that only $\sim$43\% inferences were in agreement to within a difference of 0.03 on a sample of $\sim$1,700 \textit{Kepler} red giants. To understand the origin of these differences, we analyzed a few of these stars using independent techniques such as the Monte Carlo Markov Chain method and Echelle diagrams. Through our analysis, we discovered that these alternate techniques are supportive of the neural-net inferences. We also demonstrate that the network can be used to yield estimates of coupling strength and period spacing in stars with structural discontinuities. Our findings suggest that the rate of decline in the coupling strength in the red-giant branch is greater than previously believed. These results are in closer agreement with calculations of stellar-evolution models than prior estimates, further underscoring the remarkable success of stellar-evolution theory and computation.  Additionally, we show that the uncertainty in measuring period-spacing increases rapidly with diminishing coupling strength. 
\end{abstract}

\section{Introduction}

Seismology is the most powerful tool with which to probe the interior structure of stars and understand their evolution. Missions such as \textit{CoRoT} \citep{Baglin2006a}, \textit{Kepler} \citep{Borucki2004,Borucki2010}, and TESS \citep{Ricker2015} have detected stellar pulsations in hundreds of thousands of stars, providing valuable observations. More than twenty thousand of these pulsators are red giants \citep{Mosser2010a,Yu_2018}. In the terminal stages of evolution, solar-like stars become red-giants where Hydrogen in the outer shell is depleted and core Helium becomes concentrated. Oscillations in red giants are caused by the turbulent motion of gas in the external convection zone.

Seismology has enhanced our understanding of rotation profiles in red giants and clumps \citep{Beck2012,six_stars,Di_Mauro_2016}, possible evidence for the prevalence of strong magnetic fields in the core \citep{Fuller423} and allowed for the seismic estimation of magnetic fields \citep{Li2022}. Combining asteroseismology with spectroscopic information of red giants enables us to carry out galactic archaeology \citep{10.1007/978-3-319-59315-9_5, galactic_arch}. It also sheds light on the long standing question of core-envelope angular momentum transport in evolved stars \citep{doi:10.1146/annurev-astro-091918-104359}.

To analyse observations of the \textit{Kepler} lightcurves, we follow the usual practice of transforming them into power spectra. These power spectra dominantly exhibit two kinds of resonant frequencies in solar-like stars (a) pressure driven modes (\textit{p}-modes) which are typically spaced linearly, and (b) mixed modes, whose restoring force is pressure in the envelope and buoyancy in the core. These modes propagate both in the core and envelope but decay in the evanescent zone. The modes provide us the opportunity to analyze all three zones in red giants.

Mixed modes in solar-like stars follow a highly non-linear pattern. The identification and interpretation of mixed modes is particularly challenging. With exponentially growing datasets from ongoing and future missions such as PLATO \citep{Rauer2014}, traditional fitting methods \citep{bayesian_mcmc_2009,bayesian_mcmc_2011,bayesian_mcmc_2014} are simply computationally challenging. While PLATO is primarily designed for main-sequence and subgiant stars, it is anticipated that there will be red giants in the observed field. With the potential to observe over a million light curves, PLATO could potentially capture more than 200,000 red giants if the statistics are similar to those of the Kepler mission. Although there are a few semi-automated methods \citep{Vrard2016,Gehan2018,kallinger2019release}, they still require some amount of visual inspection. It is therefore crucial to develop a fully automated, accurate and computationally cheap tool with which to analyse these stars. 

Deep learning has been very successful in identifying complex patterns in wide range of problems \citep{10.1145/3065386,Fawzi2022,Luna2020}. It has also shown success in asteroseismology in scenarios such as classifying stars \citep{10.1093/mnras/stac1515} and identifying red giants \citep{Hon2019}, measuring seismic parameters \citep{Dhanpal_2022} and measuring stellar parameters \citep{10.1093/mnras/stw1621}. In \cite{Dhanpal_2022}, we have designed an algorithm to discern pulsations of red giants from noise and measure seismic parameters that broadly describe the structure of the core and envelope of a solar-like star. Hereafter, we refer \cite{Dhanpal_2022} as D22. 

Mixed modes in red-giants can be used to probe the core of the star as shown in \cite{Bedding2011,Beck2012}. In the last paper D22, we have developed a deep learning technique to measure the asymptotic frequency spacing ($\Delta \nu$), the asymptotic period spacing ($\Delta \Pi$) of the star which is related to the core size of the star and the frequency with maximum amplitude ($\nu_{max}$) in one single step. In this paper, we extend the deep learning technique to measure the coupling strength (q). \cite{Jiang2020} have demonstrated that the asymptotic radial mode frequencies decrease as the coupling strength decreases during the evolutionary process of the star. While it has been acknowledged by \cite{Jiang2022} that a formalism to establish the relationship between coupling strength and physical quantities is lacking in cases of rapid structural variation, it has been established that the coupling strength is linked to the size of the evanescent zone in situations where the evanescent zones are thick \citep{1979PASJ...31...87S,1989nos..book.....U,Jiang_2014}. \citep{Pincon2020} have shown that the coupling strength can serve as a probe to investigate the position of the base of the convection zone. Their study also demonstrates that the progressive migration of evanescent region towards the convective zone in evolved red giants leads to decrease in coupling strength.

The automated technique developed in \citep{mixed_mode} enables the measurement of $q$ in numerous stars. However, this technique involves multiple steps, such as removing $\ell=0,2$ p-modes from the spectra, stretching the spectra, and estimating the qualitative regularity in the stretched spectra for various values of q, $\Delta \Pi$, and the offset parameter $\epsilon_{g}$. These steps require a considerable amount of time, taking at least $\sim$10s to complete. Furthermore, \cite{Ong_2023} has provided analytical evidence indicating that this approach is susceptible to overestimation due to its sole dependence on frequencies. In contrast, the trained neural network developed in this paper yields significant advantages. The end-to-end neural network is capable of inferring all seismic parameters, such as $\Delta \nu$, $\Delta \Pi$, $\nu_{max}$, and $q$, for a given star in 5 milliseconds offering a computational gain of at least $10^3$ when compared to the aforementioned automated technique. In addition, the neural network developed in this study incorporates both frequencies and amplitudes to infer the coupling strength, thereby mitigating the potential bias of overestimation.

\section{Methods and Technique}
\label{Methods_and_Technique}

To measure global seismic parameters $\Delta \nu$, $\Delta \Pi$, $\nu_{max}$ and coupling strength, we designed a neural network and trained it on synthetic spectra. Following that, we used MCMC to verify the inferences of some specific stars. In this section, we describe the neural network, the training data, the results on synthetic spectra and MCMC method used for the analysis.

\subsection{Neural Network technique}
\label{Neural_Network_technique}
Neural networks learn to execute a task by training on data. In this case, the network learns to retrieve seismic parameters \textbf{Y} from the corresponding power spectral data \textbf{X}. These networks use models with many neurons, which perform various non-linear operations to estimate the parameters. Each neuron comprises a set of tunable parameters (weights and biases) and a model typically contains millions of parameters. Neural networks learn to adjust these parameters by optimizing the error between the estimates and true labels.

Instead of directly predicting the real number associated with the parameters, we lay it out as a classification problem. To implement this, the parameter ranges of seismic parameters are divided into bins and the network estimates the probability score in each bin using a softmax function. 

The advantage of classification is that it provides a Bayesian posterior \citep{10.1162/neco.1991.3.4.461} distribution for a seismic parameter. Regression tasks typically don't offer this directly, but there are techniques like k-fold cross-validation that can estimate probability distributions \citep{10.5555/2517747}. In regression with k-fold cross-validation, the network is trained multiple times on different data subsets, increasing training time. In contrast, classification provides a probability distribution for each class with by training the network only once, eliminating the need for additional iterations or folds and significantly reducing training time. Although mixture density networks \citep{Bishop1994,10.1093/mnras/staa2853} can be used to obtain probability distributions and train the network in a single pass, it requires having a predefined number of modes and are prone to mode collapse. A thorough investigation of the advantages of different methods is beyond the scope of this paper.

In the classification framework employed, the selection of bin size holds significant importance. It is essential to strike a balance where the bin size is sufficiently small to ensure accurate parameter resolution, while also being large enough to provide an ample number of training examples within each bin for effective network training. For the current study, bin sizes of 0.1$\mu$Hz for $\Delta \nu$, 2.5s (in the range of 40-150s) and 7s (in the range of 150-500s) for $\Delta \Pi$, 2.8 $\mu$Hz for $\nu_{max}$ and 0.02 for $q$ were employed. These choices of bin sizes were approximately determined to ensure that the uncertainties in the parameters are comparable to the bin size. In the present study, we have adopted bin sizes that are representative of the typical uncertainties encountered in classical methods, such as fitting methods. However, we recognize that the uncertainties obtained by classical methods, especially in cases with low coupling constants, may be larger than typical published values for $\Delta \Pi$ as shown in section \ref{Period spacing measurements in low q stars}. In future experiments, we plan to increase the number of samples and decrease the bin size to improve precision in parameters.

We designed the neural network to estimate all four parameters at once. To carry out this task, we connected four different softmax layers representing the four seismic parameters. As shown in Figure \ref{fig:q_network_paper}, we built this network using Convolutional layers, LSTM cells, dense layers, all connected to four output softmax layers. Initially, our experimentation with the network involved solely convolutional layers. However, we have incorporated LSTM cells and increased the kernel size from 5 to 50 to improve the network's performance. Consequently, we devised the final design of the network based on these enhancements. To train this neural net, we optimize cross-entropy loss on training data using the Adam Optimizer. The network is considered trained if its performance on unseen data is as good as on training data. The trained neural network produces four probability vectors, for each power spectrum that correspond to the four parameters.

\subsection{Training data}
To train the neural network, we constructed a synthetic training dataset based on the asymptotic theory of oscillations \citep{Garcia2019,2010aste.book.....A}. This theory incorporates the physics of structure, composition gradient, and rotation in red giants. The formulation applied in the simulations is described in the Appendix A of \cite{Dhanpal_2022}. In D22, it is demonstrated that our simulations are of high quality, are realistic and cover a wide parameter space. 

\subsubsection{Modeling frequencies}

Within the predictions of asymptotic theory, dipole mixed-mode frequencies in solar-like stars are given by an implicit equation \ref{eq:mixedmodes}. Although not in its current form, the equation was originally developed in \cite{1989nos..book.....U}. It was further improved in \cite{Mosser_2012,Mosser_2015,Farnir_2021,Lindsay_2022} and \cite{Ong_2023}. In this study, we adopted the formalism presented in \cite{Mosser_2015} to facilitate a direct comparison between our neural network inferences and the values provided by the same formulation.

\begin{equation}
    \tan \pi \frac{\nu-\nu_p}{\Delta\nu} = q \tan \frac{\pi}{\Delta \Pi} \left(\frac{1}{\nu}-\frac{1}{\nu_g}\right),\hspace{0.2cm} \label{eq:mixedmodes}
\end{equation}

where $\nu_p$ are the frequencies of pure \textit{p} modes. In order to model general oscillation pattern of \textit{p} modes in red giant stars, we have employed the equation 1 given in \cite{Mosser_2012} for $\nu_p$,

\begin{equation}
    \frac{\nu_{p;n,\ell}}{\Delta \nu} = n_p + \frac{\ell}{2}+ \epsilon_p(\Delta \nu) - d_{0\ell}(\Delta \nu) + \frac{\alpha_{\ell}}{2}\left(n_p-\frac{\nu_{max}}{\Delta \nu}\right)^2,
    \label{eq:nu_p}
\end{equation}
where $\Delta \nu$ is the \textit{large-frequency separation}, which gives the mean-frequency separation between two successive radial modes, $\epsilon_p(\Delta \nu)$ is the \textit{offset parameter}, $d_{0\ell}$ the \textit{small-frequency separation}, and $\alpha_{\ell}$ the degree-dependent gradient $\alpha_{\ell} = \left(d\log \Delta \nu/dn\right)_{\ell}$.

In the equation \ref{eq:mixedmodes}, $\nu_g$ is given by, $\nu_{g}^{-1}$=$(n+\epsilon_{g})\Delta \Pi$ where $\epsilon_g$ is the offset parameter corresponding to g-modes. According to \cite{Ong_2023} and \cite{Lindsay_2022}, the relationship between the frequencies $\nu_g$ and pure g-mode frequencies ($\nu_{g,\text{pure}}$), is given by $\nu_{g}^{-1} = \nu_{g,\text{pure}}^{-1} - \Delta\Pi/2$.

To determine the frequencies of the mixed modes, we solve the implicit equation \ref{eq:mixedmodes}. By calculating the intersections between the left and right sides of the equation, we identify the mixed mode frequencies within a range of approximately 1.2$\Delta \nu$ for each pure $\ell=1$ p mode. In this process, we remove any duplicate frequencies that are within a resolution of 4 years of each other. Additional details are described in \cite{othman_benomar_2023_8296459}:external/ARMM/solver\_mm.cpp.

Pure \textit{p} mode frequencies dominantly oscillate in the envelope and pure \textit{g} mode are restricted to the core region. In contrast, mixed modes oscillate in both the core and envelope, while decaying in the evanescent zone between the core and envelope. We predominantly observe $\ell=1$ modes to be mixed modes. The coupling strength is inversely related to the size of the evanescent zone \citep{10.1093/pasj/psw104}. The transmission factor \textit{T} of the mixed mode from the \textit{g}-mode to the \textit{p}-mode cavity is related to the coupling strength, which is given by,
\begin{equation}
    T^2 = \frac{4q}{(1+q)^2}.
    \label{eq:transmission_factor}
\end{equation}
The transmission factor of the mixed mode is proportional to $\exp (-\int \kappa dr)$.  

The wave-number $\kappa$ in the evanescent zone is described by
\begin{equation}
    \label{eq:wave_number}
    \kappa = \frac{\sqrt{(\hat{S}_1^2 - \omega^2)(\omega^2 - \hat{N}_{BV}^2)}}{c\omega},
\end{equation}
where $\hat{S}_1$ and $\hat{N}_{BV}$ are the modified versions of the Lamb and Brunt-Vaisala frequencies, respectively \citep{2006PASJ...58..893T}. 

The gradients of Lamb and Brunt-Vaisala frequencies are proportional to the density contrast between the core and envelope. As the star evolves, the thickness of intermediate evanescent zone increases, causing the coupling strength to decrease further.

\begin{figure*}[!ht]
    \figurenum{1}
    \centering
    \includegraphics[width=190mm]{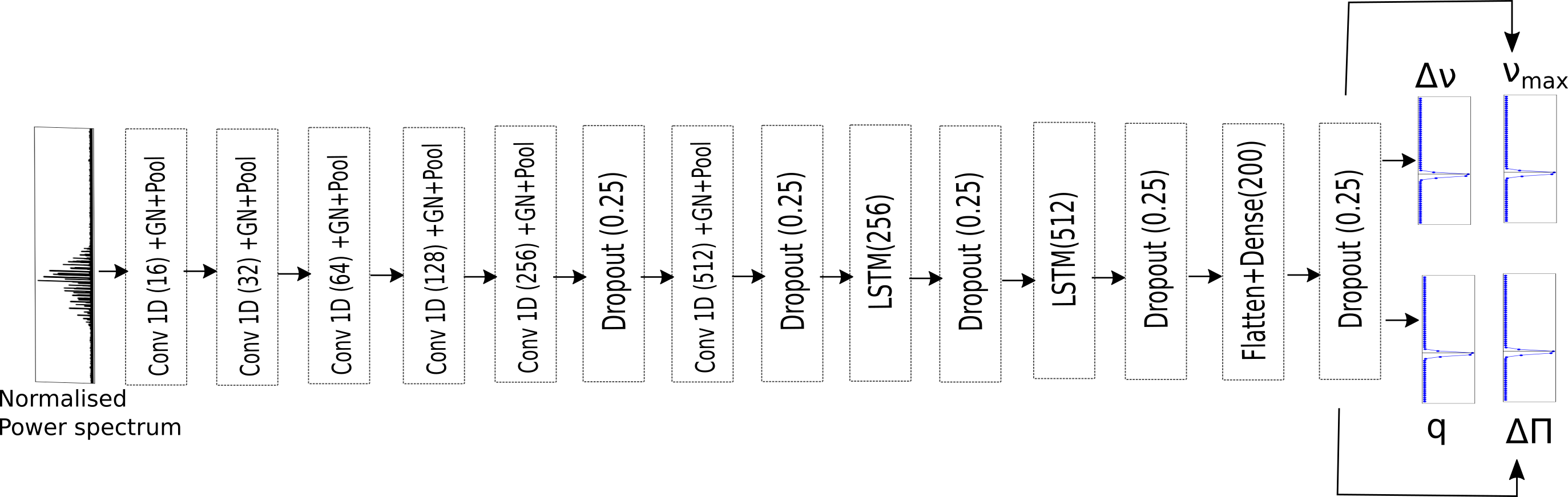}
    \label{fig:q_network_paper}
    \caption{ Architecture of neural network used to infer seismic parameters. The network takes in the normalized power spectrum as input and outputs probability distributions of $\Delta \Pi$, $\Delta \nu$, $\nu_{max}$ and $q$. The neural network comprises 1D Convolutional layers, LSTM cells and a dense layer. We used dropout layers to prevent over-fitting.}
\end{figure*}

\subsubsection{Modeling the power spectrum}
\label{Modeling the power spectrum}

The stellar power spectrum comprises the oscillation spectrum $S(\nu,\Theta_{S})$ and the noise profile $N(\nu,\Theta_{N})$. The power spectrum model \textit{M}($\nu,\Theta$) is computed using asymptotic theory, at characteristic frequency $\nu$ which is obtained as follows,

\begin{equation}
    M(\nu,\Theta) = S(\nu,\Theta_{S}) + N(\nu,\Theta_{N}).
\end{equation}

where $\Theta_{S}$ and $\Theta_{N}$ comprise the parameters of model and noise respectively.

The oscillation power spectrum is a sum of Lorentzians with heights $H(n,\ell,m)$ centered around $\nu(n,\ell,m)$ with widths $\Gamma(n,\ell,m)$. The frequencies $\nu(n,\ell,m)$ are modeled using $\ell=0,2,3$ \textit{p}-modes and dipole mixed modes given in equations \ref{eq:nu_p} and \ref{eq:mixedmodes} respectively. The oscillation spectrum model therefore is given by,

\begin{equation}
\label{eq:osc_spec_model}
    S(\nu,\Theta_{S})=\sum_{n} \sum_{\ell=0}^{\ell=3} \sum_{m=-\ell}^{m=\ell} \frac{H(n,\ell,m)}{1+4({\frac{\nu - \nu(n,\ell,m)}{\Gamma(n,\ell,m)}})^2}
\end{equation}

The heights $H(n,\ell,m)$ are modeled using the following equation,

\begin{equation}
    H(n,\ell,m) = r_{\ell,m}(\iota)V(\ell)A(n)
\end{equation}

where $V(\ell)$ is the mode visibility, $A(n)$ depends on radial order, $r_{\ell,m}(\iota)$ the relative amplitude of the mode, which depends on the inclination angle $\iota$. The visibility function is influenced by both the limb-darkening function, which varies depending on the type of star, and the measurement technique employed. For example, in a star with depressed dipole modes,  V($\ell$=1) tends to be small. 

In addition to considering the visibility factor of $\ell=1$ modes, the heights and widths of the mixed modes relative to $\ell=0$ modes are given by the following expressions:

\begin{equation}
    \Gamma_1 (\nu) = (1-\zeta(\nu))\Gamma_0(\nu);\hspace{0.2cm}A_1^2 (\nu) = (1-\zeta(\nu))A_0^2(\nu)
\end{equation}

where $\Gamma_1$ and $A_1(\nu)$ are width and amplitudes of the mixed modes respectively. In these expressions, $\Gamma_0$ and $A_0$ correspond to the width and amplitudes of the $\ell=0$ modes, which are derived from a few templates using redgiant and subgiant stars, as described in section A.1.3 of \cite{Dhanpal_2022}. And $\zeta$ is given by

\begin{equation}
    \zeta(\nu) = \left[1+\frac{\nu^2\Delta \Pi}{q\hspace{0.05cm}\Delta \nu } \frac{\cos^2\pi \frac{1}{\Delta \Pi} \left(\frac{1}{\nu} - \frac{1}{\nu_g}\right)}{\cos^2\pi \frac{\nu-\nu_p}{\Delta \nu}}\right]^{-1},
\end{equation}

In the case of a mode trapped as a p-dominated mixed mode, $\zeta$ tends to approach 0. Conversely, for a mode trapped as a g-dominated mixed mode, $\zeta$ tends to approach 1.

The noise profile is essentially due to the convective motions at the surface of the stars. The noise model is constructed as the sum of a low-frequency (lf) Harvey profile  and a high-frequency (hf) Harvey profile .  
\begin{equation}
\label{eq:osc_noise_model}
    N(\nu,\Theta_{N})=\frac{H_{\mathrm{lf}}}{1+(\tau_{\mathrm{lf}} \nu)^{p_{\mathrm{lf}}}} + \frac{H_{\mathrm{hf}}}{1+(\tau_{\mathrm{hf}} \nu)^{p_{\mathrm{hf}}}} + N_0.
\end{equation}

where H is the characteristic granulation amplitude, $\tau$ is the characteristic timescale of granulation, $p$ is the characteristic power law and $N_0$ is the white noise level.

\subsubsection{Description of Dataset}

The asymptotic theory described thus far has been implemented in \cite{othman_benomar_2023_8296459}, which generates spectra for a given set of seismic parameters. Using this theory, we built two large training datasets comprising 1 million high-frequency red-giant branch (RGB) oscillators, along with 4 million samples of low-frequency RGB and clump oscillators. The distributions of these two datasets are shown in Table \ref{tab:dataset_table}. While generating these datasets, we used a uniform random distribution to sample $\Delta\nu$, \textit{q}, $\Delta \Pi$ and most of the parameters, as shown in Table \ref{tab:dataset_table}. However, we used an isotropic distribution ($P(\iota) \propto \sin\iota$) to sample the inclination angle, and the distribution of Kepler red giants to sample the signal-to-noise ratio (SNR) and observation time.

To create these datasets, we have treated $\Delta \nu$, $\Delta \Pi$ and $q$ to be independent parameters. Within this dataset, certain combinations of seismic parameters utilized for generating spectra may lack a corresponding theoretical model for stars. Examples of such combinations include $\Delta \nu$ of 18$\mu$Hz, $\Delta \Pi$ of 41s, q of 0.5 etc., among others. Nonetheless, incorporating these examples in the training process aids the neural network to understand these seismic parameters by thoroughly investigating the diverse patterns present in the data.

\begin{deluxetable*}{cccccc}
\tablenum{1}
\tablecaption{Range of seismic parameters for the preparation of synthetic data.\label{tab:dataset_table}}
\tablewidth{0pt}
\tablehead{
\colhead{Parameter} & \colhead{High-frequency RGB stars} & \colhead{Low-frequency RGB stars} & \colhead{Sampling strategy} \\
\colhead{} & \colhead{} & \colhead{and clumps} 
}
\startdata
Range of $\Delta \nu$ & 9-19 $\mu$Hz & 1-9 $\mu$Hz & Uniform\\
Range of $\Delta \Pi$ & 40-500s & 40-500s & Uniform\\
Range of $q$ & 0-0.5 & 0-0.65 & Uniform\\
Range of $\epsilon_p$ & 0-1  & 0-1 & Uniform\\
Range of $\epsilon_g$ & 0-1  & 0-1 & Uniform\\
Range of $\alpha_{\ell}$ & 0.0-0.008  & 0.0-0.008\\
Range of Core & 0.005-2.8  & 0.005-2.8 & Uniform\\
rotation ( in $\mu$Hz) & {} & {} & {} \\
Range of Envelope & 0.005-0.4  & 0.005-0.4 & Uniform\\
rotation ( in $\mu$Hz) & {} & {} & {} \\
Range of inclination $\iota$ ( in deg) & 0-90  & 0-90 & Isotropic \\
Range of $A_{g}$  & 0.8-1.2 & 0.8-1.2 & Uniform\\
Range of $B_{g}$ & -2.2 - -1.8 & -2.2 - -1.8 & Uniform\\
Range of $C_{g}$ & 0-0.5 & 0-0.5  & Uniform\\
Range of $A_{\tau}$ & 0.8-1.2 & 0.8-1.2 & Uniform\\
Range of $B_{\tau}$ & -1.2 - -0.8 & -1.2 - -0.8 & Uniform\\
Range of $C_{\tau}$ & 0-0.5 & 0-0.5  & Uniform\\
Range of $p$ & 1.0-3.5 & 1.0-3.5 & Uniform\\
Range of $N_{0}$ & 0.001-40,000 & 0.001-40,000 & Uniform\\
Frequency range used  & 0-283$\mu$Hz  &  0-283 $\mu$Hz & {} \\
for ML training & {} & {} & {} \\
Range of Observation & 9-1460 days & 9-1460 days & Kepler \\
time & {} & {} & {} \\
SNR distribution & 8-160 & 8-160 & Kepler\\
\enddata
\tablecomments{This table shows the range of seismic parameters that were chosen to create the synthetic dataset. The range of parameters is chosen so as to cover the space of long-cadence \textit{Kepler} red-giant stars \cite{Mosser_2015,Vrard2016,mixed_mode}.}
\end{deluxetable*}

The red giants in our training data were categorized into two groups based on their $\Delta \nu$ values: low-frequency and high-frequency RGB stars. An RGB star was classified as low frequency if its $\Delta \nu$ was less than 9$\mu$Hz, while those with $\Delta \nu$ greater than or equal to 9$\mu$Hz were classified as high frequency. This classification was heuristic, taking into account findings from \cite{Mosser_2014}, which indicated that a significant portion of clumps and several hydrogen shell-burning stars fall within the parameter range of $\Delta \nu < 9.5\mu$Hz. Therefore, we set the separation point at 9$\mu$Hz. The designation of high-frequency RGB stars for $\Delta \nu>9\mu$Hz stems from their higher $\nu_{max}$ values compared to the other class, while the remaining class was referred to as low-frequency RGB stars.

Though there are two different datasets, we trained only one neural net which works across both datasets. In order to account for the lack of prior information regarding a new star's power spectrum and for the purpose of an independent analysis, we opted to train a single neural network using both datasets collectively. The neural net takes in normalized power spectra and outputs marginal distributions of all four seismic parameters ($\Delta \nu$,$\nu_{max}$,$\Delta \Pi$,$q$)
as shown in Figure \ref{fig:q_network_paper}. 

In order to optimize the utilization of our computational resources, we generated a dataset comprising 5 million examples and trained the neural network within this constraint. We anticipate that increasing the number of samples will improve the performance, and we have plans to expand the dataset in future research. However, for the purposes of this paper, we specifically focused on generating a larger number of samples for low-frequency stars. The proximity of peaks in these stars presents a more challenging task for the machine in learning the distinctive features. Consequently, we established a 4:1 ratio between the low and high frequency portions of the dataset. It is important to acknowledge that the current ratio is heuristic, and further experimentation is required to optimize the network. However, the exploration of this aspect is beyond the scope of the present paper.

\subsection{Results on Synthetic data} 
\label{Results_on_Synthetics}

For a diverse set of 5 million simulated red-giant spectra, we train the machine to obtain probability distributions of $\Delta \nu$,$\Delta \Pi$,$\nu_{max}$ and $q$ as shown in Figure \ref{fig:q_network_paper}. Although machine can infer offset parameters $\epsilon_p$ and $\epsilon_g$, the accuracy of these parameters on our simulations is relatively low and these parameters are not in the interest of this article. Hence, we do not report them.

\begin{figure*}[!ht]
     \figurenum{2}
     \centering
     \includegraphics[width=170mm]{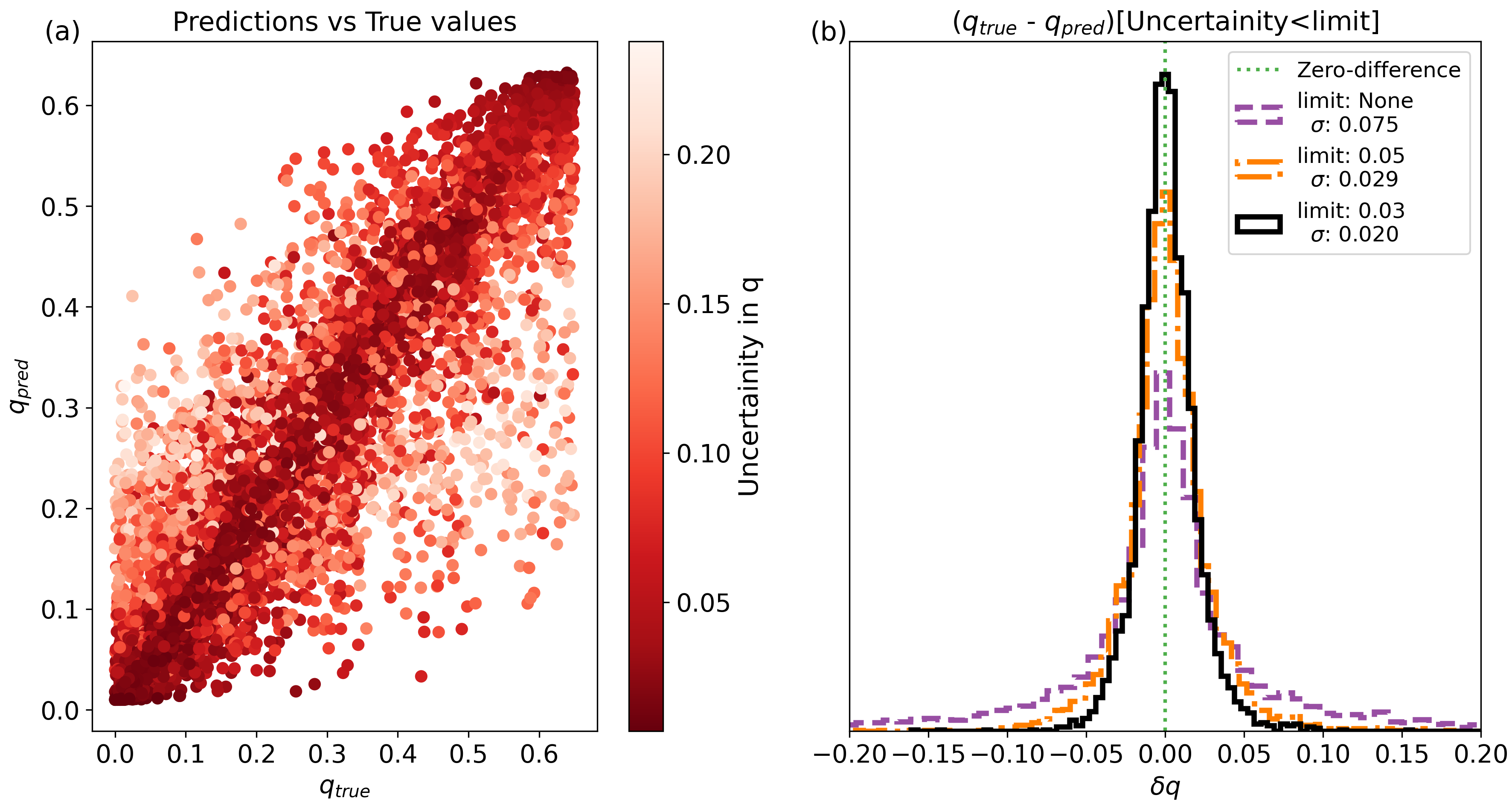}
     \caption{Results on synthetic spectra: (a) Predicted $q$ ($q_{pred}$) vs True $q$ (injected $q$) in a simulated star where the color of each point indicates uncertainty in its measurement. (b) Distribution of $\delta q$, where $\delta q = q_{true} - q_{pred}$. These distributions show error ($\delta q$) as a function of uncertainty. It shows that 67\% of predictions below uncertainty of 0.03 have $|\delta q|<0.02$. Hence, we demand that confident predictions to have uncertainty less than 0.03 in measurement of $q$.}
    \label{fig:q_results_synthetics}
\end{figure*}

Once trained, the network is evaluated over 30,000 unseen simulated spectra which span a large parameter space. We present the results of the estimates on these unseen spectra in Figure \ref{fig:q_results_synthetics}. Figure \ref{fig:q_results_synthetics}(a) plots coupling-constant predictions ($q_{pred}$) as a function of injected values ($q_{true}$). It shows that the predictions and true values are highly correlated. As the predictions uncertainty decreases, the extent of the correlation increases. Figure \ref{fig:q_results_synthetics}(b) graphs the distribution of errors in the network estimates. It shows that we can recover the coupling strength to within 0.02 for 67\% of predictions with uncertainty less than 0.03. Hence, we label the prediction with the uncertainty less than 0.03 as a confident prediction, which is a heuristic choice. 

\subsection{Markov Chain Monte Carlo (MCMC)}
\label{Monte Carlo Markov Chain (MCMC)}

Here, we briefly summarize the Markov-Chain Monte-Carlo  (MCMC)  fitting model that serves as a reference for our neural network technique. MCMC is a powerful algorithm to sample the underlying probability distribution. Compared to gradient-descent methods (MLE or MAP), MCMC is robust to local maxima and  provides accurate parameter estimates. Therefore, it serves as an ideal choice for establishing a benchmark baseline in the field of Machine Learning. In this case, we sample the two-degrees-of-freedom $\chi^2$-likelihood distribution, which is calculated using the power spectral data of the star $\textbf{y}$ and the power spectrum model $M(\nu,\Theta)$ described in section \ref{Modeling the power spectrum}.  We follow methodology described in \cite{bayesian_mcmc_2009} to fit the model to the data and get the underlying parameter distribution. Here, we briefly recall the method described in that publication.

\subsubsection{Bayesian formalism}
\label{Bayesian formalism}

To fit for seismic parameters from the given data, we use a Bayesian formalism, where we obtain the posterior distribution of the parameters given the observations $\pi(\Theta|y,M,I)$. The posterior distribution is given by 
\begin{equation}
    \begin{split}
        \pi(\Theta|y,M,I) = \prod_{i} \pi(\Theta|y_{i},M_{i},I) \\= \prod_{i} \pi(\Theta_{S}|y_{i},M_{i},I)\pi(\Theta_{N}|y_{i},M_{i},I).
    \end{split}
\end{equation}
Using Bayes' theorem, the posterior distribution $\pi(\Theta|y_{i},M_{i},I)$ is given by 
\begin{equation}
\label{eq:posterior_eq}
    \pi(\Theta|y_{i},M_{i},I) = \frac{\mathcal{L}(y_{i}|\Theta,M_{i},I) \pi(\Theta|M_{i},I)}{\pi(y|M_{i},I)},
\end{equation}
where $\mathcal{L}(y_{i}|\Theta,M_{i},I)$ is the \textit{likelihood} function, $\pi(\Theta|M_{i},I)$ is the prior of the parameters and $\pi(y|M_{i},I)$ is the evidence.

As the observed power spectrum follows  2-\textit{dof} $\chi^2$ statistics, the likelihood function $\mathcal{L}(y_{i}|\Theta,M_{i},I)$ is given by
\begin{equation}
\label{eq:likelihood_eq}
    \mathcal{L}(y_{i}|\Theta,M_{i},I) = \frac{1}{M_{i}} \exp(-\frac{y_i}{M_i}).
\end{equation}

\subsubsection{Metropolis Hastings algorithm}

For a given model $S$, we find the best fit parameters by maximizing the $\pi(\Theta|y,S,I)$. This may be achieved by maximizing the likelihood function $\pi(y|\Theta,S,I)$ or equivalently, minimizing its -$\log \pi(y|\Theta,S,I)$. However, in a Bayesian context, one needs to maximize posterior distribution as shown in equation \ref{eq:posterior_eq}.

We sample the posterior using Metropolis Hastings algorithm \citep{Metropolis1953,10.2307/2334940}. To find the best-fit parameters, we sample these posterior distributions by drawing random points from the prior and accepting at ratio $\alpha$, given by
\begin{equation}
    \alpha(\Theta_{new},\Theta_{t})= \min [1,\frac{\pi(y|\Theta_{new},S,I)}{\pi(y|\Theta_{t},S,I)}].
\end{equation}

Further details about the entire formalism can be found in \cite{bayesian_mcmc_2009}.

\section{Results on Kepler data}

After training the network on synthetic examples and building trust based on the attendant predictions, we analysed the \textit{Kepler} data. To corroborate the network performance on Kepler data, we compared the network inferences with those obtained from other methods. We have used a python package Lightkurve \citep{2018ascl.soft12013L} to extract the Kepler lightcurves and construct the power spectra.

The marginal distributions of seismic parameters obtained from network may be in varied forms and have a range of uncertainties. To compare the predictions with other methods, we only selected confident predictions, i.e., where the uncertainty in \textit{q} is $\le0.03$ and the uncertainty in $\Delta \nu$ is $\le0.1 \mu$Hz.

\subsection{Machine inferences on Kepler data}\label{Machine inferences on Kepler data}

\begin{figure*}[!ht]
    \figurenum{3}
    \centering
    \includegraphics[width=190mm]{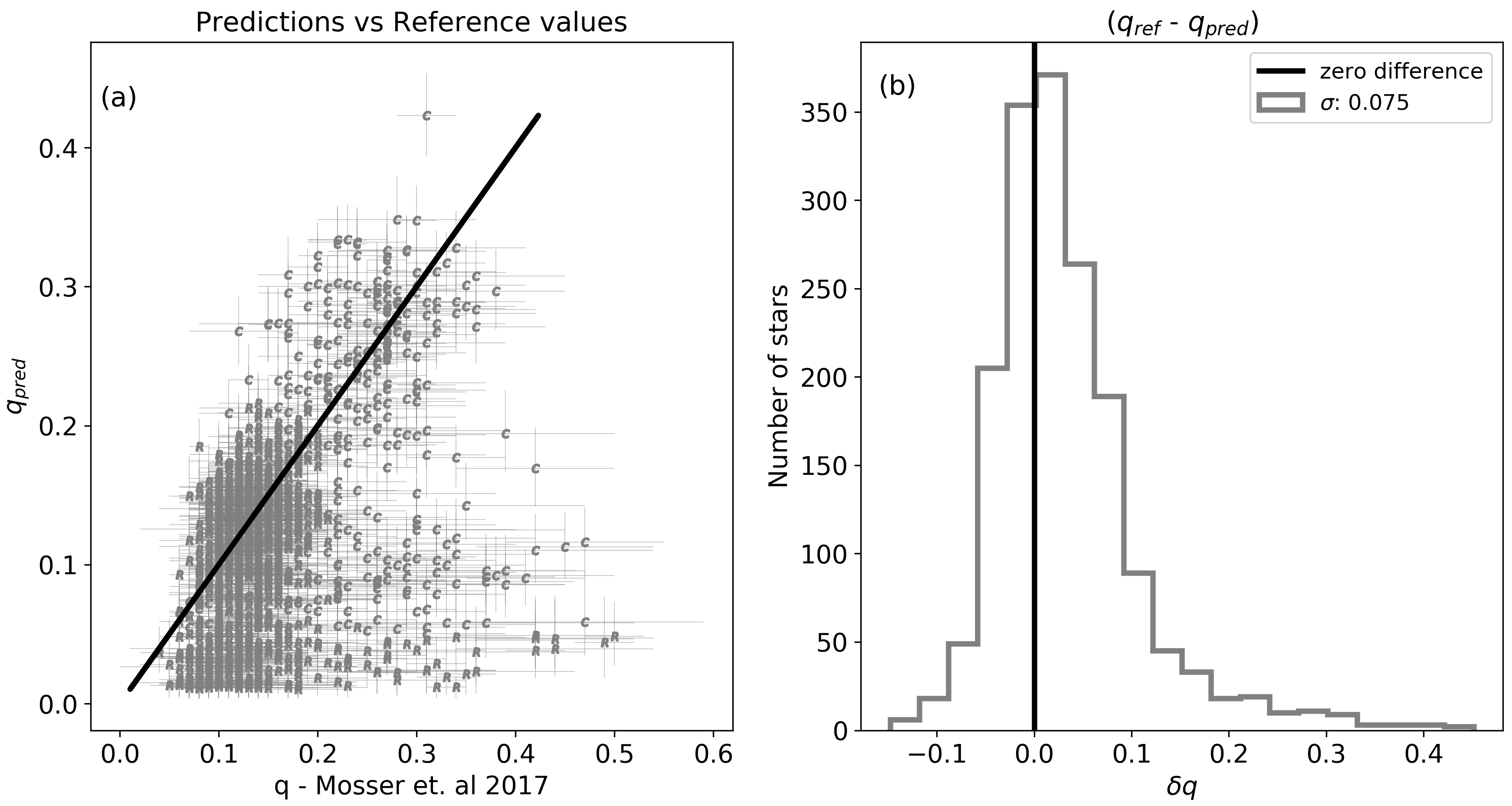}
    \caption{(a) Predicted value of $q$ by the network at each value of published $q$ in 1701 stars from \cite{mixed_mode}. 
    These 1701 stars are selected such that the estimates of both $\Delta \nu$ and $q$ are precise to within 0.1$\mu$Hz and 0.03 respectively. In this plot, the points marked with 'R' are RGB stars and the points marked with 'C' are clump stars. The grey lines associated with each point are the uncertainites and the black solid line corresponds to zero difference. The values of $\Delta \nu$ and $\Delta \Pi$ allows us to differentiate between RGB stars and clump stars \citep{Vrard2016,2013ApJ...765L..41S}. Specifically, a star is classified as a clump star if $\Delta \nu$ is less than 10$\mu$Hz and $\Delta \Pi$ is greater than 150s; otherwise it is an RGB star.(b) Distribution of difference in $q$ in all 1701 stars.} 
    \label{fig:q_results_real_data_final}
\end{figure*}

Figure~\ref{fig:q_results_real_data_final} illustrates measurements of coupling strength on 1701 stars from \cite{mixed_mode} which consists a catalogue of 5166 stars. As we are comparing only confident machine predictions, we selected stars common among the confident predictions and catalogue presented in \cite{mixed_mode}. Among 5166 stars, 53.8\% of the network's measurements exhibit lower uncertainties in comparison to those reported in \cite{mixed_mode}, and this percentage remains at 47.4\% among the subset of 3465 stars not meeting the confidence criteria for measurements.

In Figure~\ref{fig:q_results_real_data_final}(a), we have plotted the neural net's predictions ($q_{pred}$) against published values ($q_{ref}$) and figure~\ref{fig:q_results_real_data_final}(b) graphs the distribution of differences between the respective values. In 730 stars, the differences between predictions and published values are less than 0.03. In 309 stars, the published values are greater than predictions by at least 0.08. In 301 stars, the coupling strength measurements of the machine are $\le$ 0.05. As the network's measurements in these 309 stars are lower than published values, we investigated some of these stars by constructing echelle diagrams.

\begin{figure*}[!ht]
     \figurenum{4}
     \centering
     \includegraphics[width=170mm]{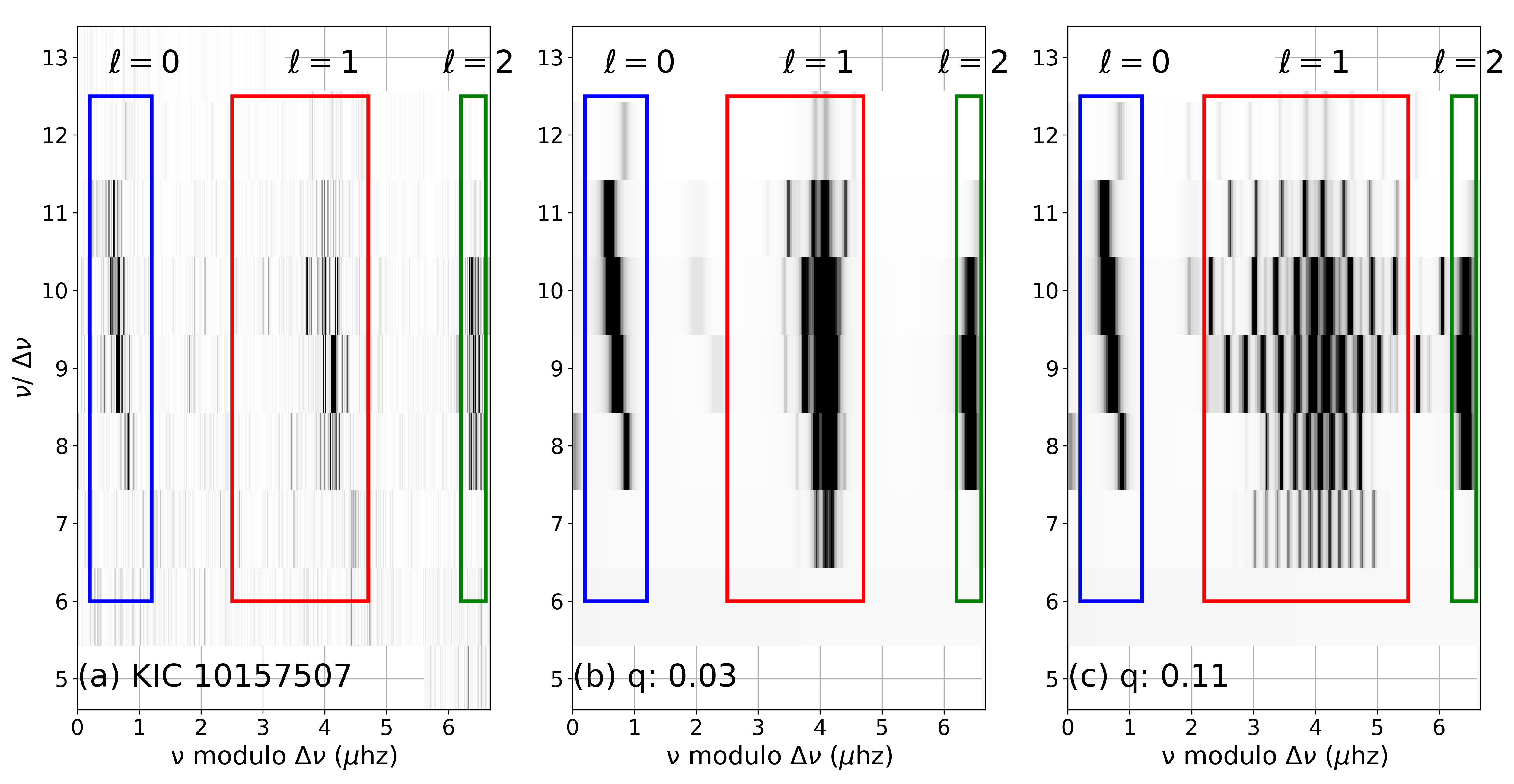}
     \caption{(a) Echelle diagram associated with KIC 10157507. (b,c)
     Echelle diagrams of typical simulated stars with $q=0.03$ and $q=0.11$. 
     In all these echelle diagrams, blue rectangles denote $\ell=0$ \textit{p}-modes, green rectangles indicate $\ell=2$ \textit{p}-modes and red rectangles mark $\ell=1$ mixed modes. The echelle diagram of KIC 10157507 is closer to the simulated echelle diagram associated with $q=0.03$ when compared with that of $q=0.11$, indicating that the star's $q\leq0.05$.}
    \label{fig:q_synthetics_merged}
\end{figure*}

Figure~\ref{fig:q_synthetics_merged} shows an example of echelle diagram analysis. For KIC 10157507, the network's measurement of $q$ is $0.03\pm0.02$ whereas \cite{mixed_mode} measured the coupling strength to be $\sim$0.11. In figures \ref{fig:q_synthetics_merged}(b) and \ref{fig:q_synthetics_merged}(c), we show synthetic echelle diagrams with injected $q=0.03$ and $q=0.11$ respectively. As $q$ increases from 0.03 to 0.11, the number of distinct and apparent $\ell=1$ mixed modes increase. In fact, there is a conspicuous difference in the amplitude of $g$-dominated modes between the two cases. When we compare the echelle diagram of KIC 10157507 shown in \ref{fig:q_synthetics_merged}(a) to synthetic analogues, the star is qualitatively closer to an echelle diagram with $q=0.03$.

\begin{figure*}[!ht]
     \figurenum{5}
     \centering
     \includegraphics[width=180mm]{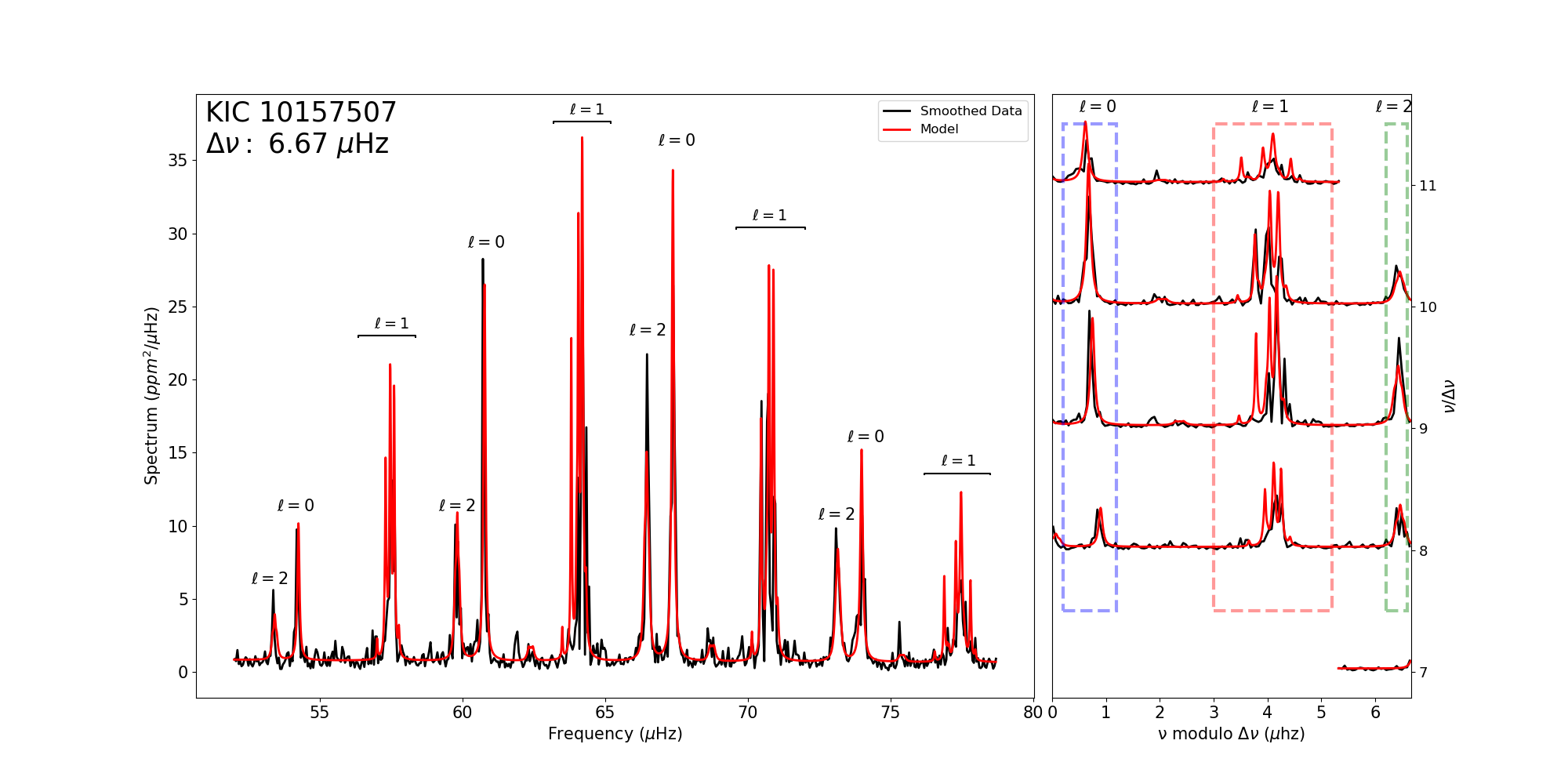}
     \caption{(\textit{left}): Comparison of best-fit model obtained from MCMC with smoothed power spectrum of KIC 10157507. We smoothed the spectra using a box-car of window size 0.05$\mu$Hz. The best-fit model was generated using $q_{\mathrm{fit}}=0.027$, $\Delta \Pi_{\mathrm{fit}}=74$s, which are median values of their respective distributions. (\textit{right}): It shows the comparison between the echelle diagram of the best-fit model and the smoothed data. These plots indicate that the best-fit model matches the observations.}
    \label{fig:best_fit_model_kic_10157507}
\end{figure*}

\begin{deluxetable}{ccccc}
\tablenum{2}
\tablecaption{Prior of different seismic parameters for analysis of KIC 10157507.\label{tab:prior_mcmc}}
\tablewidth{0pt}
\tablehead{
\colhead{Parameter} & \colhead{Prior distribution} & \colhead{Range of prior}
}
\startdata
$\Delta \Pi$ & Uniform & 40 - 150s\\
\hline
$\epsilon_{g}$ & Uniform & 0 - 1.0\\
\hline
$q$ & Uniform & 0 - 1.0\\
\hline
Inclination $\iota$ & Uniform & 0 - 90$^{\circ}$ \\
\hline
$\Omega_{core}/2\pi$ & Uniform & 0 - 2.0$\mu$Hz \\
\hline
$\Omega_{env}/2\pi$ & Uniform & 0 - 0.3$\mu$Hz \\
\hline
Visibility $V(\ell=1)$ & Uniform & 0.05 - 5.0 \\
\hline
Visibility $V(\ell=2)$ & Uniform & 0.05 - 0.8 \\
\hline
$\ell=0$ Frequency 1  & Uniform & 54.17 - 54.39 $\mu$Hz \\
\hline
$\ell=0$ Frequency 2  & Uniform & 60.66 - 60.88 $\mu$Hz \\
\hline
$\ell=0$ Frequency 3  & Uniform & 67.36 - 67.59 $\mu$Hz \\
\hline
$\ell=0$ Frequency 4  & Uniform & 73.96 - 74.18 $\mu$Hz \\
\hline
$\ell=0$ Frequency 5  & Uniform & 80.82 - 81.04 $\mu$Hz \\
\enddata
\tablecomments{Following a visual inspection of the power spectrum, we constrain the prior of $\ell=0$ frequencies to the specified ranges presented in the table.}
\end{deluxetable}

To establish additional evidence, we have also analysed this star using MCMC. We obtain the underlying posterior distributions on KIC 10157507 using MCMC with prior shown in Table \ref{tab:prior_mcmc}. The best-fit model is computed using the medians of the posterior distributions. We show comparison of the best-fit model to smoothed data in figure \ref{fig:best_fit_model_kic_10157507}. We smoothed the data using a box-car of width 0.05$\mu$Hz. Figure~\ref{fig:best_fit_model_kic_10157507}(a) shows the mode distribution of the smoothed data and the model and figure \ref{fig:best_fit_model_kic_10157507}(b) shows the comparison of the mode distribution as an echelle diagram. The echelle diagram indicates that the mode distribution matches the smoothed data across all modes ($\ell=0,1,2$) statistically.

\begin{figure*}[!ht]
     \figurenum{6}
     \centering
     \includegraphics[width=170mm]{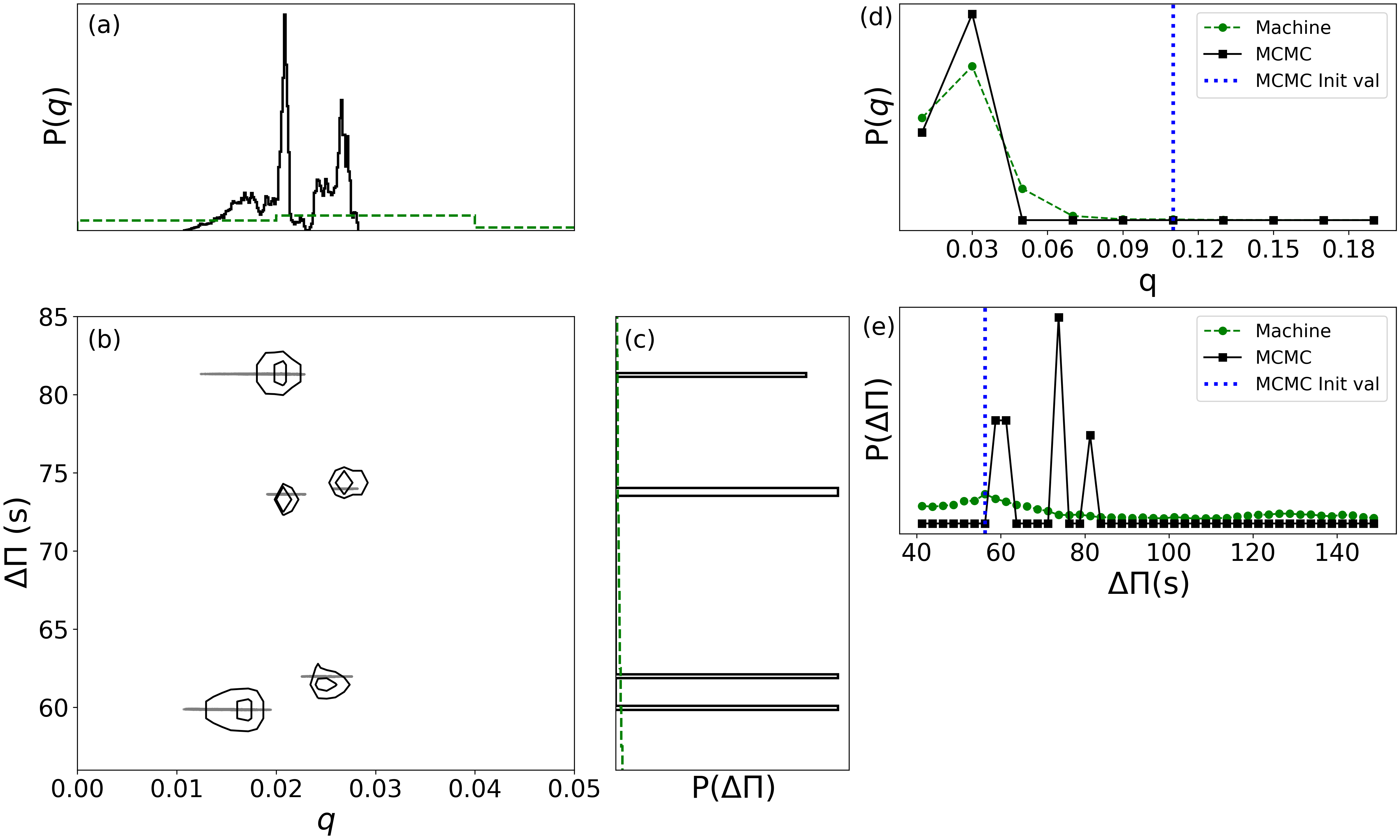}
     \caption{Panels (a), (b), and (c) display the MCMC distributions in a corner plot for KIC 10157507. Panel (b) specifically presents the MCMC distribution in the $q$-$\Delta \Pi$ plane, which includes the 68\% and 95\% confidence intervals. The marginal distributions of $q$ and $\Delta \Pi$ obtained through MCMC are shown in figures (a) and (c), respectively. The green dashed lines in both panels represent the distributions obtained by the neural network. Panel (d) compares the distributions of the coupling strength obtained from both methods, MCMC and the neural network. The blue line represents the initial value used in the MCMC run, with $q$ initialized to a value larger than the neural network's prediction to avoid bias. The agreement between the two distributions supports our conclusion. In figure (e), the distributions of the period spacing ($\Delta \Pi$) from the MCMC and neural network methods are compared. The blue line represents the MCMC initialization. The two distributions do not converge to each other. The bin sizes in panels (d) and (e) correspond to the bin sizes of the neural network, for a better comparison with the network's outputs.}
     \label{fig:q_dp_comparison_distribution}
\end{figure*}

\begin{figure*}[!ht]
     \figurenum{7}
     \centering
     \includegraphics[width=170mm]{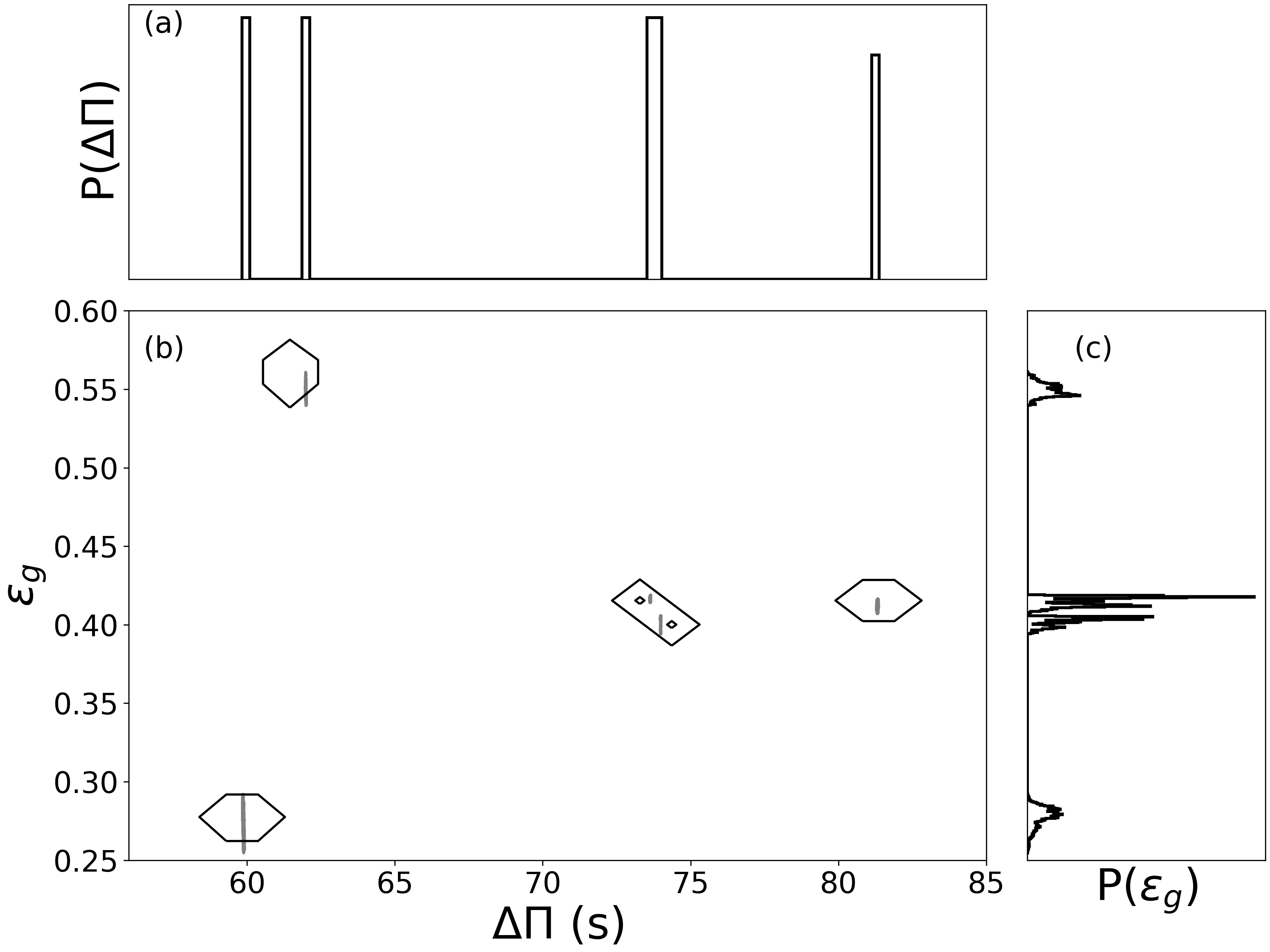}
     \caption{Panels (a), (b), and (c) display the MCMC posterior distributions of $\Delta \Pi$ and $\epsilon_{g}$ in a corner plot for KIC 10157507. Panel (b) presents the MCMC distribution in the $\Delta \Pi$-$\epsilon_{g}$ plane, which includes the 68\% and 95\% confidence intervals. The individual marginal distributions of $\Delta \Pi$ and $\epsilon_{g}$ obtained through MCMC are depicted in panels (a) and (c), respectively.}
     \label{fig:dp_epg_comparison_distribution}
\end{figure*}

\begin{figure*}[!ht]
    \figurenum{8}
    \centering
    \includegraphics[width=170mm]{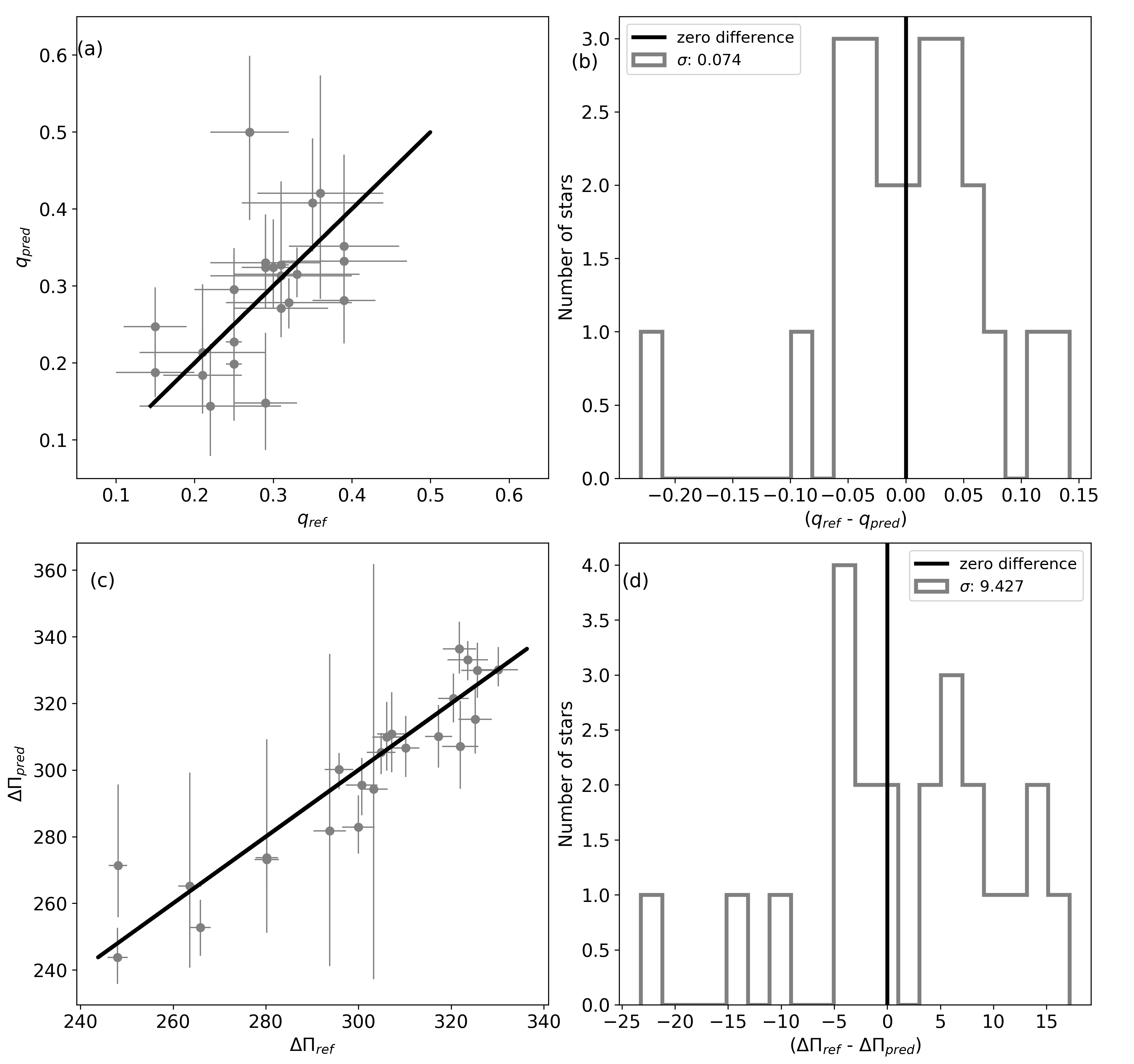}
    \caption{(a) Estimated values of $q$ by the network ($q_{pred}$) corresponding to each reported value of $q$ in 23 stars from \cite{Vrard2022} ($q_{ref}$) (b) Distribution of the differences between the estimated and reported values in these stars.(c) Estimated values of $\Delta \Pi$ by the network ($\Delta \Pi_{pred}$) corresponding to each reported value of $\Delta \Pi$ ($\Delta \Pi_{ref}$) in these stars. (d) Distribution of difference in these two independent measurements of $\Delta \Pi$ in these stars. The red lines associated in panels (a) and (c) with each point are the 1-$\sigma$ uncertainties.} 
    \label{fig:q_results_glitch_stars_edited}
\end{figure*}

We present the mixed-mode parameter distributions of $q$ and $\Delta \Pi$ in figure \ref{fig:q_dp_comparison_distribution}. In the context of a specific neural network, the training data serves as a prior. The neural network developed here outputs probability distributions that can be seen as analogous to marginal posterior distributions obtained through MCMC sampling. As the neural network outputs are Bayesian \citep{10.1162/neco.1991.3.4.461}, comparing them to MCMC is a like-to-like comparison. However, the uncertainties in their output distributions also depend on network complexity and the number of training samples. Therefore, contrasting these distributions to MCMC posterior distributions can provide validation for the uncertainties and distributions we have obtained. Figure \ref{fig:q_dp_comparison_distribution}(a) suggest that the distributions of coupling strength obtained by the neural net and  MCMC are in statistical agreement, i.e., to within an errorbar. The current network cannot accurately replicate the MCMC posterior distribution due to its large bin size. But both these methods affirm that $q<0.05$ for KIC 10157507, demonstrating that some red-giants may have a measurable low coupling strength. 

\subsection{Inferences on clumps with glitches}
\label{Inferences on red giants with glitches}

Figure~\ref{fig:q_results_glitch_stars_edited} showcases the measurements of the coupling strength and period spacing on 23 stars from \cite{Vrard2022} that exhibit structural discontinuities\footnote{We have considered the stars given in \href{https://www.nature.com/articles/s41467-022-34986-z/tables/1}{https://www.nature.com/articles/s41467-022-34986-z/tables/1}}. In Figure~\ref{fig:q_results_glitch_stars_edited}(a), we plot the neural network's estimates of the coupling strength ($q_{pred}$) against the reported values ($q_{ref}$), while Figure~\ref{fig:q_results_glitch_stars_edited}(b) displays the distribution of differences between the respective values. Similarly, in Figure~\ref{fig:q_results_glitch_stars_edited}(c), we plot the neural network's estimates of the period spacing ($\Delta \Pi_{pred}$) against the reported values ($\Delta \Pi_{ref}$), and Figure~\ref{fig:q_results_glitch_stars_edited}(d) illustrates the distribution of differences between these values.

It can be observed that the uncertainties in the neural network's estimates of $q$ are larger than 0.03 for the majority of the stars, which does not meet our criterion for confident predictions. Despite the lack of precision, the medians of the distributions obtained from the neural network exhibit a strong correlation with the values reported in \cite{Vrard2022}. The neural network measurements of $q$ and the measurements of \cite{Vrard2022} have a 55\% correlation, while the neural network measurements of $\Delta \Pi$ and the measurements of \cite{Vrard2022} exhibit a 92\% correlation. Out of the 23 stars, 14 stars agree with the measurements of \cite{Vrard2022} within 0.05 for $q$, and 17 stars agree within 10s for $\Delta \Pi$. The choice of a threshold of 0.05 for $q$ and 10 seconds for $\Delta \Pi$ for statistical agreement was made based on the approximate mean uncertainty for these stars for those respective parameters. It is worth noting that larger uncertainties are expected since the training data does not encompass such variations.

\subsection{Period-spacing measurements in low \textit{q} stars}
\label{Period spacing measurements in low q stars}

\begin{figure*}
     \figurenum{9}
     \centering
     \includegraphics[width=170mm]{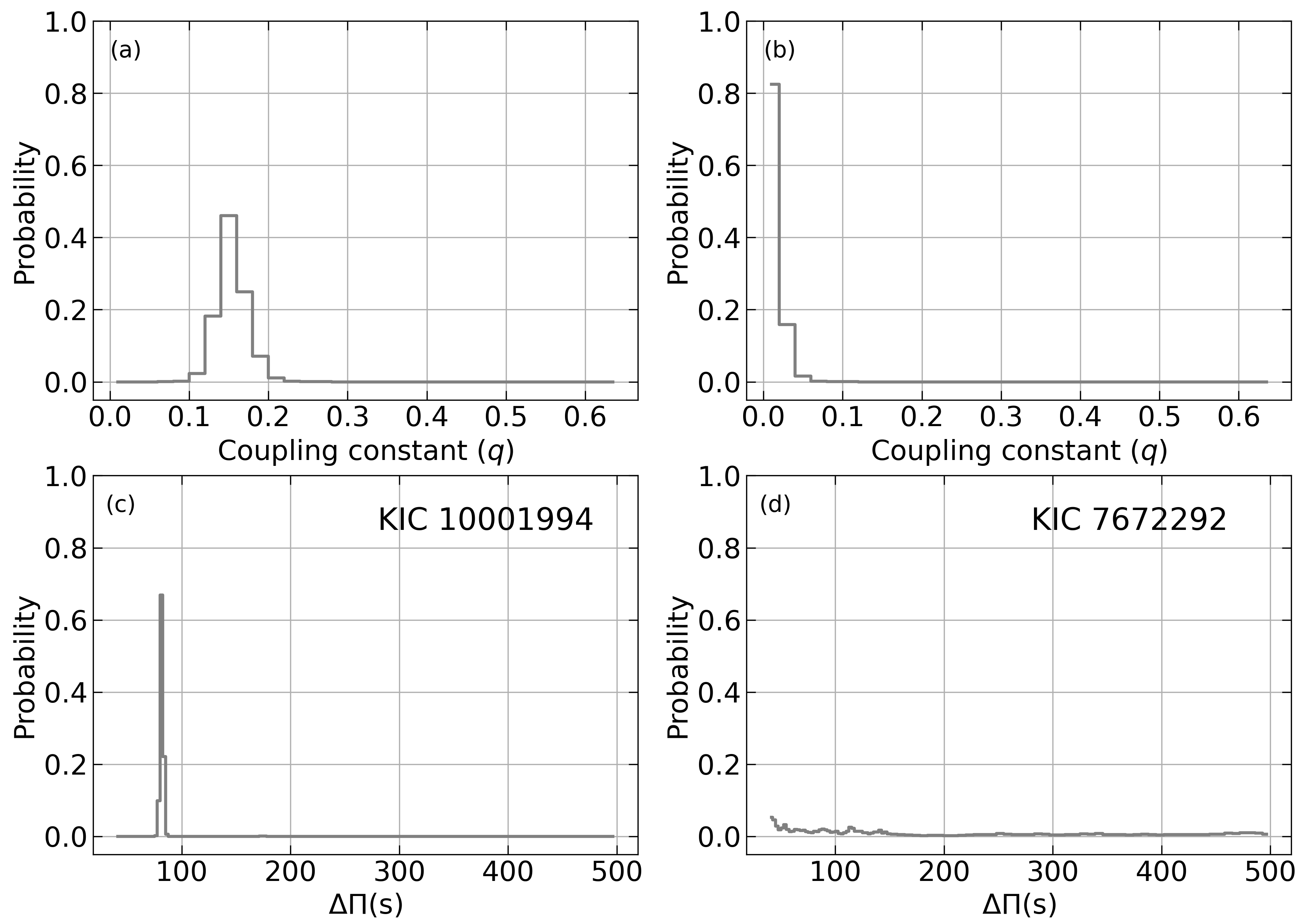}
     \caption{Probability distributions of $q$ and $\Delta \Pi$ in two stars KIC 10001994 and KIC 7672292. In the first star, $q = 0.15 \pm 0.02$ and $p_{max}$ of the $\Delta \Pi$ distribution is $\sim 0.7$. In the second star, $q = 0.01 \pm 0.01$ and $p_{max}$ of the $\Delta \Pi$ distribution is $\sim 0.05$. It is evident that $\Delta \Pi$ is well-constrained when $q$ is larger (KIC 10001994, $q$=0.15) and the distribution is flat when $q$ is smaller (KIC 7672292, $q$=0.01).}.
    \label{fig:prob_dist_dp_q_7672292}
\end{figure*}

In D22, we showed that the machine-learning model is successful at measuring $\Delta \Pi$ in red giants across all evolutionary stages. We show measurements of $\Delta \Pi$ in two stars with low coupling strength in figures \ref{fig:q_dp_comparison_distribution}(b) and \ref{fig:prob_dist_dp_q_7672292}.

As displayed in figures \ref{fig:q_dp_comparison_distribution}(a) and \ref{fig:prob_dist_dp_q_7672292}(b), KIC 10157507 and KIC 7272292 have coupling strengths that are less than 0.05. Figures \ref{fig:q_dp_comparison_distribution}(b) and \ref{fig:prob_dist_dp_q_7672292}(d) indicate that the network's $\Delta \Pi$ distributions are not exhibiting any prominent peaks, and that precise estimation of $\Delta \Pi$ may not be possible. We established this also using MCMC, as seen in \ref{fig:q_dp_comparison_distribution}(b). We find that $\Delta \Pi$ does not converge to a single value, instead possessing three distinct peaks in its distribution, thereby suggesting the possibility that $\Delta \Pi$ cannot be constrained in this star. It implies that  $q$ can be measured reliably even if $\Delta\Pi$ cannot be constrained.

While the neural net is unable to output precise estimates of $\Delta \Pi$ in stars with low \textit{q}, it is able to return accurate estimates in stars with high \textit{q}. For example, in KIC 10001994, \textit{q} is well above 0.1 and the measurement of $\Delta \Pi$ is precise as presented in figure \ref{fig:prob_dist_dp_q_7672292}(c). 

\begin{figure}[!ht]
     \figurenum{10}
     \centering
     \includegraphics[width=80mm]{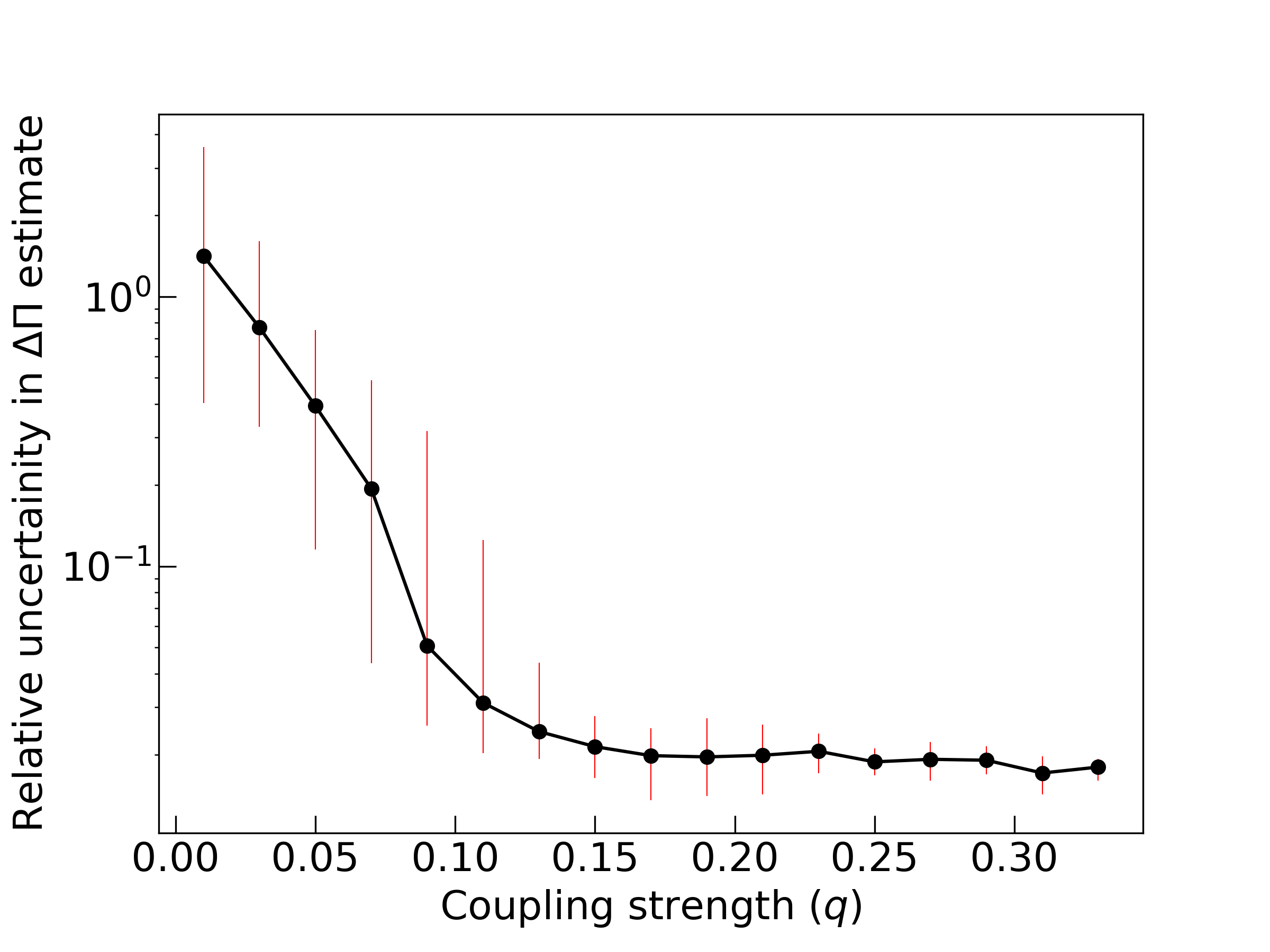}
     \caption{Relative uncertainty in  $\Delta \Pi$'s measurement as a function of coupling strength ($q$). The black dot and red lines associated with each point indicate median and 1-$\sigma$ interval of relative uncertainty distribution in $\Delta \Pi$ predictions for each value of $q$. As $q$ increases from 0.01 to 0.15, the relative uncertainty in $\Delta \Pi$ decreases from $\sim$1 to $\sim$0.02.}
    \label{fig:dp_confidence_q}
\end{figure}

From the three cases described so far, it is seen that the precision in measuring $\Delta \Pi$ is highly correlated with the coupling strength. To illustrate this, we studied the uncertainty in inferring $\Delta \Pi$ as a function of \textit{q}. We present these results in figure \ref{fig:dp_confidence_q}, where we plot the variation of relative uncertainty at each level of \textit{q}. As \textit{q} increases, the mean uncertainty in $\Delta \Pi$ predictions improves. This study establishes that, for a measurement with relative uncertainty of $\Delta \Pi$ less than 0.2, q needs to be greater than $\sim$0.05.

The correlation between the precision of $\Delta \Pi$ measurement and coupling strength \textit{q} in figure \ref{fig:dp_confidence_q} may also be understood from theory. At low values of the coupling strength, the transmission factor is low and the amplitudes of \textit{g}-dominated mixed modes are tiny. And since the transmission factor is low, modes encounter significant decay in the evanescent region, implying that the Brunt-Vaisala cavity cannot be accurately probed. This further indicates that $\Delta \Pi$ cannot be measured precisely, as seen in our observations.

The likelihood function for the 2-degree-of-freedom $\chi^2$ distribution may have a weak convexity, with several local peaks arising from the fact that it is a high dimensional parameter space, incorporating parameters pertaining to period spacing, amplitudes, widths, and noise profile. It may not lead to a unimodal distribution in the MCMC fits. As a result, we observe the bimodality in Figure \ref{fig:q_dp_comparison_distribution}(a). However, this does not mean that $q$ is unconstrained. We verify convergence through analysis of the acceptance ratio of samples, falling within the range of 0.2-0.25, as per \cite{bayesian_mcmc_2009}. Additionally, we assess convergence through the stability of likelihood and posterior distributions across multiple parallel chains. As seen in Figure \ref{fig:q_dp_comparison_distribution}(a), $q$ is constrained to lie within the range 0.01-0.03, as opposed to $\Delta \Pi$, for which the uncertainty is large in its distribution. Each peak in the bimodal distribution of $q$ leads to multiple solutions for $\Delta \Pi$, which further proves that $\Delta \Pi$ cannot be constrained for these low values of the coupling constant. Since $\Delta \Pi$ cannot be constrained, $\epsilon_{g}$ cannot be constrained either, as shown in Figure \ref{fig:dp_epg_comparison_distribution}.

The strong correlation between $\Delta\Pi$ and $\epsilon_g$ in their joint posterior distribution is expected, as they are both important in the determination of g-mode frequencies. However, the neural network struggles to accurately infer $\epsilon_{g}$. In order to demonstrate that the uncertainty in $\Delta \Pi$ in stars with low coupling constant is not an artifact of determining $\epsilon_{g}$, we present similar findings using MCMC. Figure \ref{fig:dp_epg_comparison_distribution}(b) provides evidence in KIC 10157507 that identical values of $\epsilon_{g}$ around 0.4 can yield divergent solutions for $\Delta \Pi$. This substantiates the assertion that, even with the successful constraining of $\epsilon_{g}$, the capacity to accurately infer $\Delta \Pi$ in these particular stars remains limited.

\subsection{Coupling strength as a function of large separation}
\label{Coupling strength as a function of large separation}

\begin{figure*}[!ht]
     \figurenum{11}
     \centering
     \includegraphics[width=170mm]{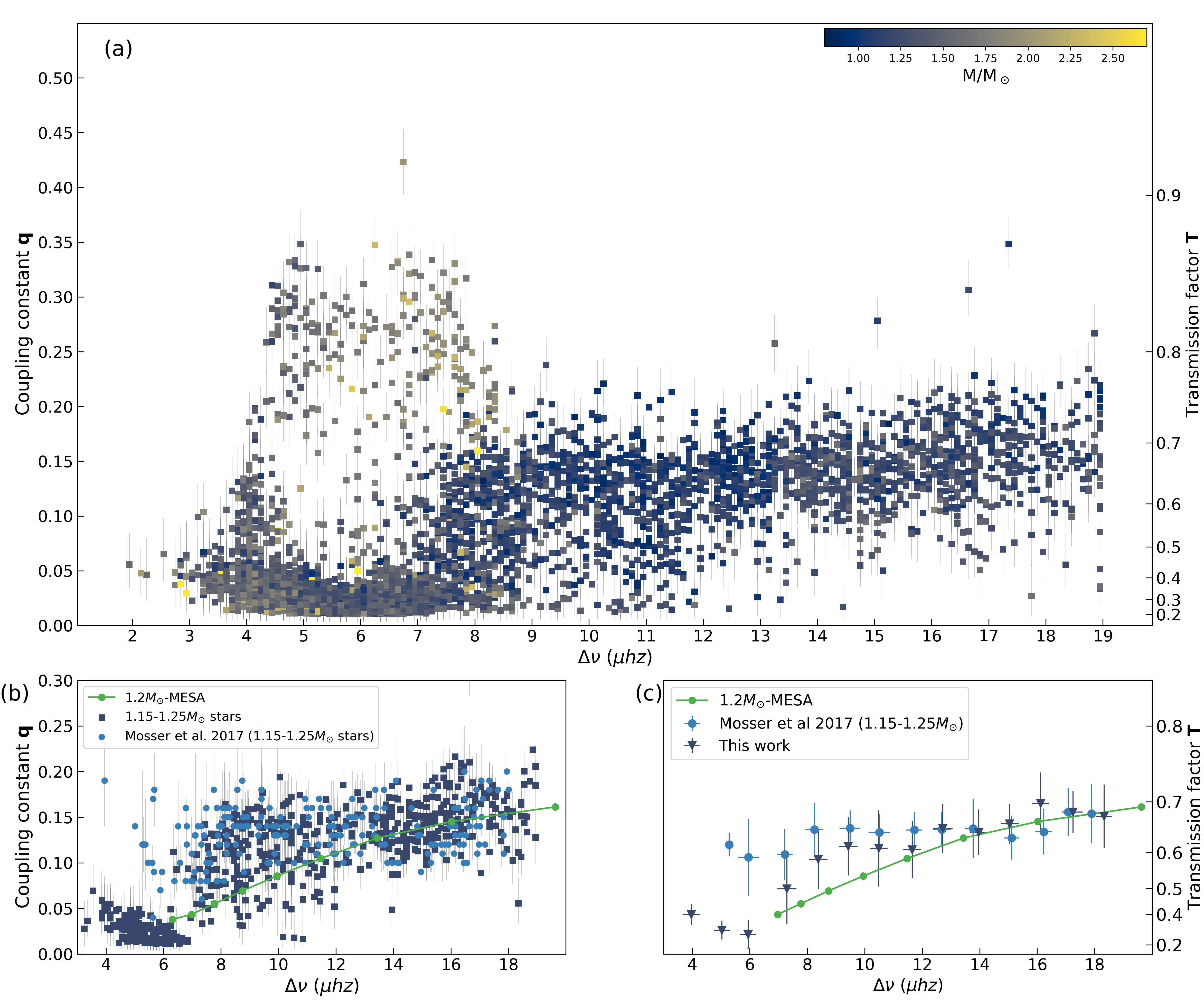}
     \caption{(a)Network-predicted $q$ vs. $\Delta \nu$ predictions in 5443 stars. The color of each point represents the ratio of the mass of the star to the solar mass, as calculated by a scaling relation \citep{2012sse..book.....K,1991ApJ...368..599B,Mathur2012}. Low-mass RGB stars and high-mass clumps split into two branches around $\Delta \nu$ of $\sim$8$\mu$Hz in this distribution and reunify in later stage of evolution i.e., around $\Delta \nu$ of $\sim$4$\mu$Hz. (b) This plot compares the $q$-$\Delta \nu$ distributions obtained from our work and the catalogue of \cite{mixed_mode} for hydrogen shell-burning red giants within the mass range of 1.15-1.25 $M_{\odot}$, along with the theoretical model of a 1.2 $M_{\odot}$ star. The blue circular points represent data from the catalogue, while the dark square points represent data from our work. Each point is accompanied by grey lines that indicate the 1-$\sigma$ errors in the value of $q$. (c) This plot presents a comparison between the distributions of ($\langle \Delta \nu \rangle$,$\langle q \rangle$) obtained from our work (dark inverted triangles) and the catalogue of \cite{mixed_mode} (blue circles). The vertical lines indicate the standard deviations of the $q$ predictions within each $\Delta \nu$ bin. Additionally, the green line represents the theoretical relationship between $q$ and $\Delta \nu$ derived from MESA.}
    \label{fig:q_dnu}
\end{figure*}

We show the results of the correlation between coupling strength and large separation in figure \ref{fig:q_dnu}. For this study, we selected a sample of 5443 stars out of $\sim$21000 red giants in the Kepler catalog. As the neural net was able to measure both $q$ and $\Delta \nu$ precisely within 0.03 and 0.1$\mu$Hz respectively for these 5443 stars, we have selected this sample. It may be seen that there are two branches in the distribution of $q-\Delta \nu$ which split at $\Delta \nu \sim 8 \mu$Hz. The first branch has low mass RGB stars, with $\Delta \nu$ spanning from 3$\mu$Hz to 19$\mu$Hz and $q \le 0.21$ approximately. The second branch comprises high-mass He-burning clump stars with $q \ge 0.15$ and $\Delta \nu<10\mu$Hz. These two branches reunite in their later evolved stages when $\Delta \nu \sim 4 \mu$Hz. The study also shows a low density of stars in the phase space of $\Delta \nu$ $\in$ [5-6.5]$\mu$Hz and $q\in$ [0.08,0.2]. This observation can potentially be attributed to a combination of the reduction in the coupling strength in the hydrogen shell-burning red giant branch \citep{Jiang2020} and higher coupling strengths in He-burning stars, leading to an absence of stars in this particular region.   

To compare observations with theory, we have simulated a 1.2$M_{\odot}$ star using the stellar evolution code \citep[MESA][]{MESA2011, MESA2013, MESA2015, MESA2018} using the inlist of \citealt{takata_masao_2019_2594045} and calculated $q$ at each evolutionary step. Coupling strength $q$ is computed based on the asymptotic theory, which is originally developed in \citet{10.1093/pasj/psw104} and extended in \citet{10.5281/zenodo.1874121}. It is given by,

\begin{equation}
    \label{eq:q_X}
    q(X) = \frac{1-\sqrt{1-e^{-2\pi X}}}{1+\sqrt{1-e^{-2\pi X}}},
\end{equation}
where the quantity $X$ is given by the following equation
\begin{equation}
    \label{eq:X_kappa}
    X \propto \int_{\varepsilon} \kappa \mathrm{dr} + X_{R}.
\end{equation}
In equation~(\ref{eq:X_kappa}), $\kappa$ is calculated using equation (\ref{eq:wave_number}). Note that $X_{R}$ is a correction term that is important when the evanescent region is thin.

While equation \ref{eq:q_X} does not account for stars with structural glitches, as demonstrated in \citep{Jiang2022}, we assumed that the stars follow asymptotic theory. The simulated spectra are based on the assumption of asymptotic theory, assuming no variation in $q$ and $\Delta \Pi$ as a function of frequency. Therefore, it is meaningful to compare the inferences derived from the network to the theoretical simulations without any discontinuities. We have taken care to eliminate spikes from the model structure and calculate the coupling strength accordingly. Consequently, the calculation of $q$ in this study differs from that of \citep{Jiang2022}. 

We calculate $q$ using equation (\ref{eq:q_X}) for a 1.2$M_{\odot}$ star and select 664 hydrogen shell-burning red giants with masses in the range 1.15-1.25 $M_{\odot}$ from our sample to compare with the theory. Figure~\ref{fig:q_dnu}(b,c) shows $q$ as a function of $\Delta \nu$ for the set of 664 red giants against the theoretical calculation of $q$. In our simulation, as the star evolves, $\Delta \nu$ undergoes a decrease from $\sim$20$\mu$Hz to $\sim$6$\mu$Hz that is correlated with $q$ dropping from $\sim$0.16 to $\sim$0.04. The figure indicates that observations match the theory statistically within 1$\sigma$ intervals across evolutionary stages \footnote{The measurements on clump stars have not been compared to theory as theory is not known and out of scope of this paper}. 

Furthermore, in Section \ref{Inferences on red giants with glitches}, we have demonstrated that even in the presence of glitches, our neural network is able to recover the correct value of $q$ with only increased uncertainty. Since the stars under consideration do not exhibit larger uncertainties, and most of these stars have $\Delta\nu > 6 \mu$Hz unlike \cite{Jiang2022}, it is possible that they do not possess significant discontinuities. 

\cite{Ong_2023}  claim that the measurements of $q$ reported in \cite{mixed_mode} may have been potentially overestimated. To explore the possible connection with this claim, we plot Figures \ref{fig:q_dnu}(b,c) by analyzing a subset of 200 hydrogen shell-burning RGB stars from \cite{mixed_mode} with masses ranging from 1.15 to 1.25 $M_{\odot}$. Notably, as the red giants evolve from approximately 18 $\mu$Hz to 6 $\mu$Hz, the average value of $q$ in our study decreases from around 0.15 to 0.02, whereas the average value of $q$ in \cite{mixed_mode} decreases from approximately 0.15 to 0.10. This analysis highlights that the overestimation observed in \cite{mixed_mode} increases from 0.02 at 11 $\mu$Hz to 0.08 at 6 $\mu$Hz in our investigation.

It has been suggested by \cite{Ong_2023} that if the sampling discriminant, $D_{\text{samp}}=\frac{2}{\pi}qN_1$,  
where $N_1 = \Delta \nu/\nu_{max}^2 \Delta \Pi$, satisfies $D_{\text{samp}}<1$, then it can lead to an overestimation of the coupling strength. In Section \ref{Machine inferences on Kepler data}, we demonstrate that out of $\sim$1700 stars, there is a close agreement in the coupling strength measurements within a difference of 0.03 for 730 stars between the two methods (our work and \cite{mixed_mode}). However, among these 730 stars, 375 stars satisfy the condition $D_{\text{samp}}<1$ from the measurements obtained by both from our network and \cite{mixed_mode} but the network doesn't perceive any significant overestimation. Moreover, these stars do not suffer from the large uncertainty issue illustrated in section \ref{Period spacing measurements in low q stars} and the estimates of $\Delta \Pi$ are reliable. Therefore, the explanation provided in \cite{Ong_2023} may not be sufficient to account for the observed overestimation of the coupling strength in other stars as perceived by the network.

\section{Conclusion}
We present a deep-learning algorithm to measure the coupling strength in red-giant stars. The neural network was trained on a large library of 5 million synthetic examples which resemble observations. These synthetic simulated spectra were computed using an asymptotic theory of oscillations. We demonstrated that our neural network is calibrated and extracts the coupling strength accurately on synthetic spectra (Section \ref{Results_on_Synthetics}). 

In a dataset of 5166 stars, 53.8\% of the network's measurements demonstrated reduced uncertainties compared to those reported in \cite{mixed_mode}. A comparison of our confident estimates on 1701 stars to the method from \citep{mixed_mode} showed that 730 stars (42.9\%) among these are in agreement with \cite{mixed_mode} to within a difference in the coupling strength of $<0.03$. In 309 stars (18.2\%), the neural network infers these measurements to be lower than previously known measurements by $\sim$0.08.

We observed that our neural network finds that $\sim$17.5\% of these 1701 stars have a low coupling strength, i.e., $q\le0.05$. We analysed an example KIC 10157507 from this sample of stars using MCMC to validate the measurement of this star. The parameter distribution of $q$ obtained from MCMC agrees with the network-predicted distribution. For this star, the MCMC distribution of $\Delta \Pi$ does not converge to one particular value which has also been observed in the neural net's $\Delta \Pi$ distribution. Therefore, \textit{q} is $\sim$ 0.03 atleast statistically and $\Delta \Pi$ can't be determined for this star.

By analyzing a sample of 23 clump stars with structural discontinuities \citep{Vrard2022}, we observed that the neural network is capable of estimating the coupling strength within a range of 0.05 for 14 stars, and the period separation ($\Delta \Pi$) within 10 seconds for 17 stars. The results of this study illustrate the potential use of neural networks for obtaining estimates of $q$ and $\Delta \Pi$, even in the presence of structural glitches. 

One advantage of the deep-learning method is that $q$ can be measured even if $\Delta \Pi$ can't be constrained. In addition, we observed that the inference of $\Delta \Pi$ depends on the value of coupling strength (c.f. Figure \ref{fig:dp_confidence_q}). The $\Delta \Pi$ measurements obtained using the neural network are not precise when $q<0.05$. Our study shows that uncertainty in measurement decreases with the value of the coupling strength. These findings may also be explained as the coupling strength is proportional to the square of transmission factor of the mixed mode originating from \textit{g}-mode cavity. As the transmission increases with $q$, the amplitudes of $g$-dominated mixed modes improve which help in constraining $\Delta \Pi$.

Our deep learning technique, which incorporates both amplitudes and frequencies, may outperform other methods for constraining the coupling strength in stars. However, a comprehensive analysis is needed to fully understand the differences between our method and others, which is beyond the scope of this paper.

The correlation between the coupling strength ($q$) and large frequency separation ($\Delta \nu$) in the red-giant branch (RGB) is well-established \citep{mixed_mode}. Remarkably, the neural network, trained on synthetic spectra devoid of artificial correlations, independently identifies a strong correlation between $q$ and $\Delta \nu$ in Kepler RGB stars. This observation serves as validation for both the neural network and the synthetic spectra developed in this study. While this relationship has been previously observed by \citep{mixed_mode}, the network's inferences have a slightly steeper slope in the correlation.

Theoretical calculations of a 1.2$M_{\odot}$ star show that the change in coupling strength as a function of the large separation $\Delta \nu$ agrees with observations of 1.2$M_{\odot}$ stars to within 1$\sigma$ at various evolutionary stages. This study explains the marginally stronger correlation discovered by the neural net. Our results indicate that the measurements of \cite{mixed_mode} have been overestimated. \cite{Ong_2023} proposes an insightful explanation for the observed overestimation. Among the $\sim$1700 stars in the sample, $\sim$510 stars have been overestimated by at least 0.05, and they satisfy the condition of sampling discriminant ($D_\text{samp}=\frac{2q\Delta\nu}{\pi \nu_{max}^2 \Delta \Pi}$) values less than 1 as specified in \cite{Ong_2023}. However, there are stars that meet this condition, yet our network does not exhibit the same overestimation.

The network is able to extract the coupling strength from $\sim$1,000 stars in under $\sim$5 seconds, enabling ensemble asteroseismology on vast data sets. The current network can only provide reliable estimates of the coupling strength for a quarter of the entire Kepler red giants. It may encounter challenges in other samples due to low SNR, structural glitches, or inherent limitations of the network. In future work, we will improve this fraction and expand it to infer all global seismic parameters, such as core and envelope rotation rates, and inclination angle, by combining this method with Monte-Carlo-based techniques.

\textbf{Acknowledgment:} S.D. acknowledges SERB, DST, Government of India, CII and Intel Technology India Pvt. Ltd. for the Prime minister's fellowship and facilitating research. M.T. is partially supported by JSPS KAKENHI Grant Number 18K03695. This research was supported in part by a generous donation (from the Murty Trust) aimed at enabling advances in astrophysics through the use of machine learning. Murty Trust, an initiative of the Murty Foundation, is a not-for-profit organisation dedicated to the preservation and celebration of culture, science, and knowledge systems born out of India. The Murty Trust is headed by Mrs. Sudha Murty and Mr. Rohan Murty. All computations are performed on Intel\textsuperscript{\textregistered} Xeon\textsuperscript{\textregistered} Platinum 8280 CPU. We thank Dhiraj D. Kalamkar, Intel Technology India Pvt. Ltd. for suggestions which helped optimize the neural-network training. We thank Prof. Benoit Mosser for the useful discussions. We thank the anonymous referee for their invaluable suggestions, which enhanced the quality of the paper. This paper includes data collected by the \textit{Kepler} mission and obtained from the MAST data archive at the Space Telescope Science Institute (STScI). Funding for the \textit{Kepler} mission is provided by the NASA Science Mission Directorate. STScI is operated by the Association of Universities for Research in Astronomy, Inc., under NASA contract NAS 5–26555. This research made use of Lightkurve, a Python package for \textit{Kepler} and TESS data analysis (Lightkurve Collaboration, 2018). 

\software{
    GYRE\citep{GYRE2013,GYRE2017,GYRE2020},  
    Lightkurve \citep{2018ascl.soft12013L}, 
    MESA \citep{MESA2011,MESA2013,MESA2015,MESA2018}
}

\bibliography{scibib}{}
\bibliographystyle{aasjournal}

\end{document}